\documentclass{PoS}
\usepackage[outercaption]{sidecap}

\title{A walk through the world of chiral dynamics}

\ShortTitle{A walk through the world of chiral dynamics}

\author{\speaker{Ulf-G. Mei{\ss}ner}\thanks{Work supported in part by
         DFG, EU and BMBF.}\\
        Helmholtz-Institut f\"ur Strahlen- und Kernphysik and Bethe
        Center for Theoretical Physics
        Universit\"at Bonn,  D-53115 Bonn, Germany \\
        Institute for Advanced Simulation, Institut f\"{u}r Kernphysik
        and J\"ulich Center for Hadron Physics,
        Forschungszentrum J\"{u}lich,
        D-52425 J\"{u}lich, Germany \\
        E-mail: \email{meissner@hiskp.uni-bonn.de}}

\abstract{Hadron-hadron scattering lengths are fine probes of 
          our understanding of nonperturbative QCD. I discuss the
          status of a variety of scattering processes sensitive to the
          spontaneous and explicit chiral symmetry breaking of QCD,
          such as pion-pion, pion-kaon, pion-nucleon,
          antikaon-nucleon and Goldstone boson scattering off $D$-mesons.
          The fruitful interplay of theory, experiment and lattice QCD
          is emphasized. I point out what has to be done in these fields
          to gain further insight into these fundamental parameters.}

\FullConference{The 7th International Workshop on Chiral Dynamics,\\
		August 6 -10, 2012\\
		Jefferson Lab, Newport News, Virginia, USA}

\begin{document}

\section{Why hadron-hadron scattering?}

In 1966, Weinberg considered pion scattering off hadrons
using current algebra techniques~\cite{Weinberg:1966kf}.
For  pion scattering on a target with mass $m_t$ and isospin $T_t$,
the corresponding scattering lengths for total isospin $T$ read
\begin{equation}
a_T = -\frac{L}{1+ M_\pi/m_t}\, \left[T(T+1) -T_t(T_t+1) -2 \right]~,
\end{equation}
with $M_\pi$ the charged pion mass. For pion scattering on a 
pion [``the more complicated case''], he found
\begin{equation}
a_0 = \frac{7}{4}L ~, ~~~ a_2 =
-\frac{1}{2}L~, ~~~
L = \frac{g_V^2 M_\pi}{8\pi F_\pi^2} \simeq 0.1 \, M_\pi^{-1}~,
\end{equation}
with $g_V \simeq 1$ the vector coupling and $F_\pi \simeq 92\,$MeV the
weak pion decay constant. The predictions were on one side amazing 
due to their extreme simplicity and on the other side surprising,
as one believed that scattering lengths should be of the order
of 1~fm, the typical hadronic length scale - and not much smaller
as given by these equations. The physics behind this suppression
is well understood - the Goldstone boson nature of the pions requires
decoupling from any given field as external momenta go to zero (in the
limit of vanishing quark / pion masses). Corrections to these 
predictions can be worked out consistently in chiral perturbation theory
(CHPT) and variants thereof, as will be discussed in the next sections. 
 CHPT is the effective field theory (EFT) of the Standard Model at low energies
and allows one to systematically explore the consequences of the
spontaneous and explicit chiral symmetry breaking in QCD. Given the aims
and scope of this meeting, it is therefore natural to ask: what have 
we learned  since the seminal work of Weinberg? In the following, 
I will give a very personal answer to this question and hopefully 
convince the reader
that hadron-hadron scattering is a fine tool to gain insight into the
strong interactions in the nonperturbative regime.

\section{Chiral symmetry and the essence of chiral perturbation theory}

This section serves as a warm up -- as CHPT and extensions thereof will be
used to analyze various hadron-hadron scattering processes, a few remarks are
in order.
As already stated, CHPT  explores the consequences of chiral symmetry breaking
in QCD. As in any effective field theory, a power counting based on the scale
separation lets one organize the string of contributions to a given matrix
element or Greens function. The power of chiral symmetry manifests itself in
the relations between many processes. A particularly nice example of this are
the next-to-leading order low-energy constants (LECs) of the chiral effective
pion-nucleon Lagrangian, commonly denoted as $c_i$. These can e.g. be
determined  from data on low energy pion-nucleon scattering to some precision.
They can then be used in the chiral EFT description of the nuclear 
forces, as they
play an important role in the two-pion exchange contributions of the
two-nucleon forces and also in the leading long-range part of the 
three-nucleon force. It is also important to stress that while the number
of LECs increases with higher orders, this is not necessarily 
prolific for basic processes. E.g., elastic $\pi\pi$ scattering to
one loop features four LECs, at two loops there are only 2 
new  LECs, as all other ${\cal O}(p^6)$ local operators 
simply lead to a quark mass renormalizations of the  ${\cal O}(p^4)$ 
LECs. Further, the predictions of CHPT can  
be sharpened by combining it with other methods such as dispersion relations,
coupled-channel approaches, lattice simulations and so on. As I will show,
this can lead to incredibly precise predictions in some cases, but one
should be aware that there is no free lunch - as will be discussed in the following.

\section{Lesson 1: Pion-pion scattering}

 Elastic pion-pion scattering ($\pi\pi\to\pi\pi$) is the purest process 
in two-flavor chiral dynamics as the up and the down quark masses are
really small compared to any other hadronic mass scale. At threshold 
the scattering amplitude is given in terms of two numbers, the 
scattering lengths $a_0$ and $a_2$. Most interesting is the history
of the prediction for $a_0$: At   leading order (LO) (tree graphs in CHPT)
one has $a_0 = 0.16$ \cite{Weinberg:1966kf}. The NLO (one-loop) corrections
were worked out by Gasser and Leutwyler in 1983, $a_0 = 0.20 \pm 0.01$
\cite{Gasser:1983kx}. The fairly large correction can be understood
in terms of the strong $\pi\pi$ final-state interactions in this partial
wave. At  NNLO (2-loops), one finds $a_0 = 0.217 \pm 0.009$ 
\cite{Bijnens:1995yn} and this was considerably sharpened by 
matching the  2-loop representation to the Roy equation solution,
resulting in $a_0 = 0.220 \pm 0.005$ \cite{Colangelo:2000jc}. This is an
amazingly precise prediction for a low-energy QCD observable. Incidentally,
a similarly accurate prediction was achieved for $a_2$, 
$a_2 = -0.0444 \pm 0.0010$, but here the corrections 
to the LO result, $a_2^{\rm LO} = -0.0042$, are very small due to 
the very weak $\pi\pi$ interactions
in isospin~2. This visible improvement in accuracy compared to the two-loop
predictions was due to the inclusion of data from higher energies in the
dispersion relations with better accuracy than this is possible through the
finite number of LECs. Also, the two scattering lengths serve as subtraction
constants in the Roy equations and are thus much tighter constrained than
in the pure chiral expansion. 

\begin{SCfigure}[][t]
\includegraphics[width=0.54\textwidth]{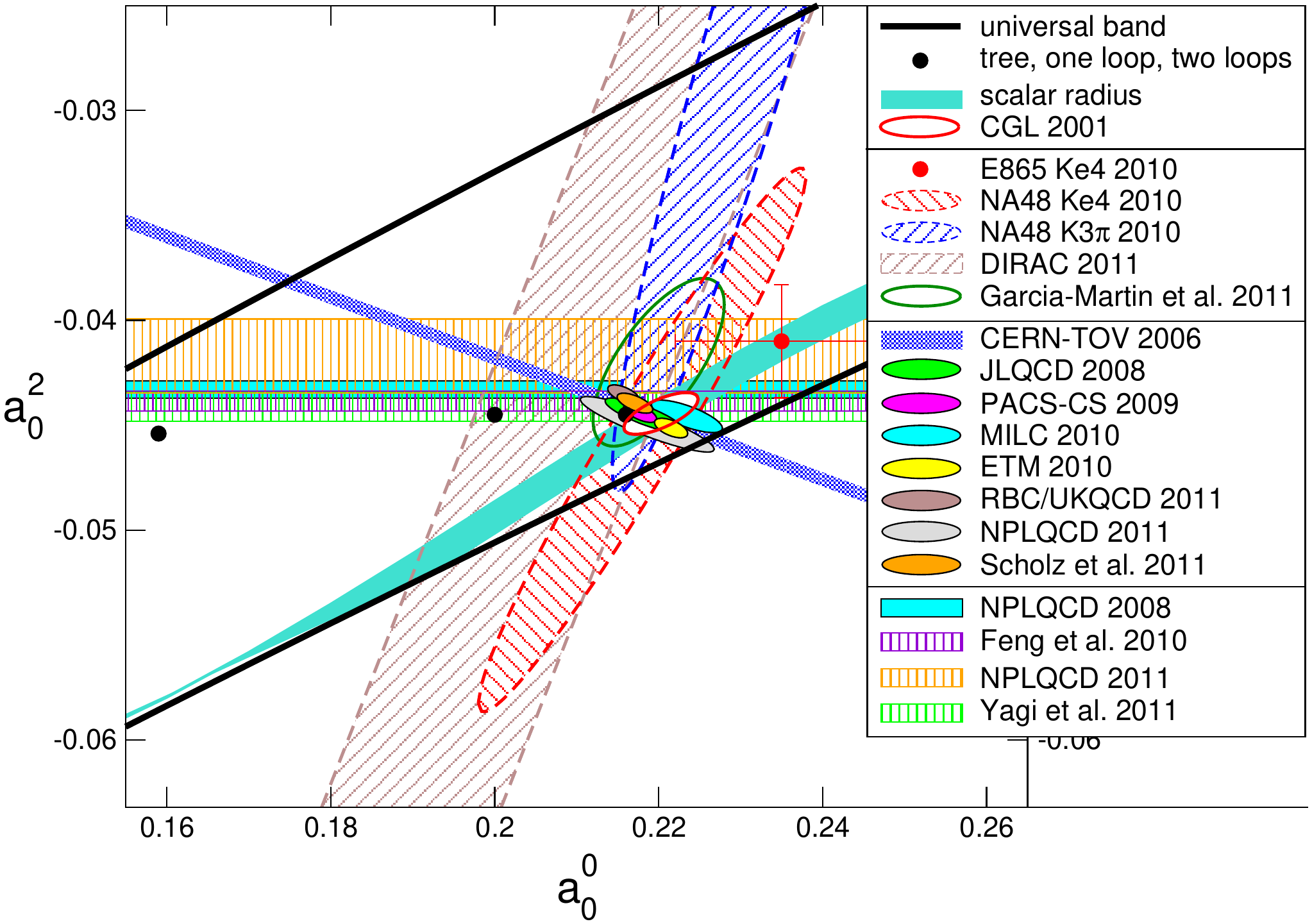}~~~~~
\caption{Theoretical predictions for the S-wave $\pi\pi$
  scattering lengths in comparison to experimental data 
 as well as direct and indirect lattice determinations. Figure
courtesy of Heiri Leutwyler \cite{Heiri}.
}
\label{fig:pipi}
\end{SCfigure}

Given such precise predictions - how about
experiment? The analysis of the $\pi\pi$ final-state interactions  in
$K_{e4}$ and the cusp in $K^0\to 3\pi^0$ decays has proven to lead to the most precise
determinations of the scattering lengths. An alternative it the measurement
of lifetime of pionium, but this is experimentally more difficult and thus
less accurate. From kaon decays using a particularly tailored non-relativistic
EFT \cite{Gasser:2011ju}, one obtains ~$a_0^0 = 0.2210\pm 
0.0047_{\rm stat}\pm 0.0040_{\rm sys}$ and $a_0^2 = -0.0429\pm 
0.0044_{\rm stat}\pm 0.0028_{\rm sys}$ \cite{Batley:2010zza}.
The pionium lifetime measurement leads to $|a_0^0-a_0^2| = 
0.2533^{+0.0080}_{-0.0078}{}^{+0.0078}_{-0.0073}$, where the first/second error
is statistical/systematic \cite{Adeva:2011tc}. The agreement with the
prediction from Ref.~\cite{Colangelo:2000jc} is stunning.  In addition, there
are direct and indirect lattice determinations of these fundamental
parameters. Here, ``direct'' refers to using the L\"uscher method 
and extracting the scattering length from the measured energy shift 
while ``indirect'' means that
the LECs $\ell_3$ and $\ell_4$ have been extracted from the pion decay constant
and mass whereas $\ell_{1,2}$ have been taken from other sources. The grand
picture is presented in Fig.~\ref{fig:pipi}  and shows a 
beautiful consistency.
This is truly  one of the finest tests of the Standard Model at low energies.
However, not all is well --  a direct lattice determination of $a_0$ is still
missing and the lattice practitioners are urged to provide this so important
number. Such a calculation is, of course, technically challenging because
of the disconnected diagrams, but time is ripe for doing it.

\section{Lesson 2: Pion-kaon scattering}

\begin{SCfigure}[][t]
\includegraphics[width=0.54\textwidth]{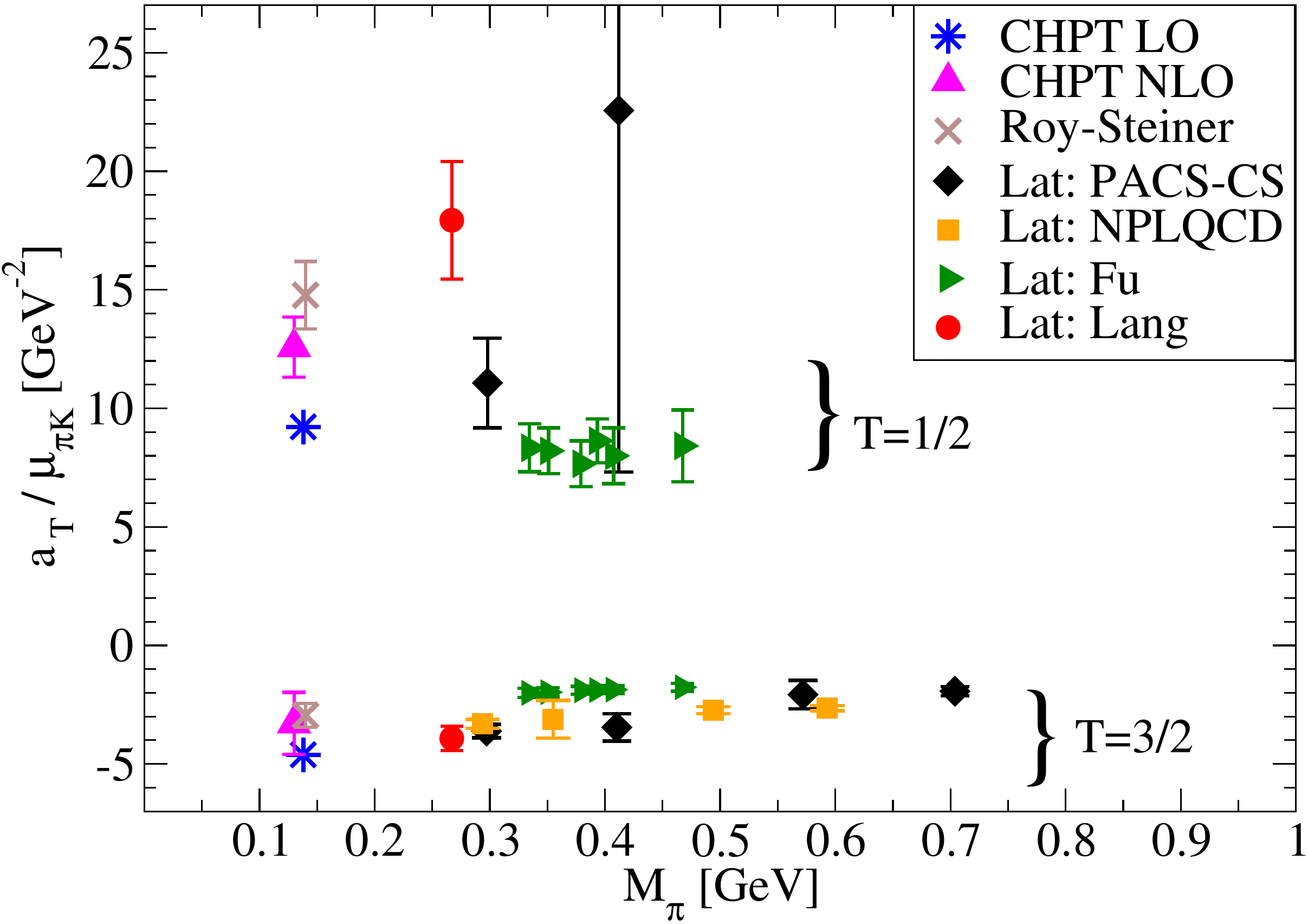}~~~~~
\caption{The S-wave $\pi K$ scattering lengths  from the lattice 
in comparison to LO and NLO CHPT and Roy-Steiner determinations. 
Data as collected in Ref.~\cite{Lang:2012sv}. For better comparison, 
the scattering lengths are normalized to the reduced mass 
$\mu_{\pi K}$ of the pion-kaon system.
}
\label{fig:Kpi}
\end{SCfigure}

The purest scattering process in chiral dynamics involving strange
quarks is elastic pion-kaon scattering, $\pi K\to\pi K$. Again, at 
threshold the scattering amplitude is given in terms of two numbers,
namely the scattering lengths $a_{1/2}$ and  $a_{3/2}$.  Before 
discussing the status of the $\pi K$ scattering lengths, I want to
address a few mysteries surrounding the $s$ quark. In standard 
three-flavor CHPT, it is treated like the $u,d$ quarks. However: 
is the strange quark really light as $m_s \sim 100~{\rm MeV}
\sim \Lambda_{\rm QCD}$? This is reflected in the expansion parameter: 
$\xi_s = {M_K^2}/{(4\pi F_\pi)^2} \simeq
0.18$, which is much bigger than its SU(2) equivalent: 
$\xi = {M_\pi^2}/{(4\pi F_\pi)^2} \simeq 0.014$.
In fact,  many predictions of SU(3) CHPT work quite well, but
there are also indications of bad convergence in some recent 
lattice calculations, see e.g. Refs.~\cite{Boyle:2007qe,Allton:2008pn}.
A possible solution to this is offered by reordering techniques,
see~\cite{Bernard:2010ex}.
Also, since many years there have been speculations that the
three-flavor condensate $\Sigma(3)$ is very much suppressed compared to its
two-flavor cousin $\Sigma(2)$. E.g. Moussallam performed a sum rule study
and found a sizeable suppression, $\Sigma (3) =  \Sigma (2) 
[1-0.54\pm 0.27]$ \cite{Moussallam:1999aq}
whereas a recent lattice study finds a more standard relation
$\Sigma (3) = \Sigma(2) [1-0.23\pm 0.39]$ \cite{Fukaya:2010na}.
Note that in both cases the uncertainties are large.
The history of the CHPT predictions for the scattering lengths
\begin{table}[t]
\begin{center}
\begin{tabular}{|l|c|c|c|c|}
\hline
 & Tree \cite{Weinberg:1966kf,Griffith:1969ph}   
 &  1-loop \cite{Bernard:1990kx,Bernard:1990kw}
 &  2-loop \cite{Bijnens:2004bu}
 & RS \cite{Buettiker:2003pp} \\
\hline
$a_0^{1/2}$ & +0.14 & $+0.18\pm 0.03$ & +0.220 [0.17 \ldots 0.225] &
    $+0.224\pm 0.022$\\
$a_0^{3/2}$ & $-0.07$ & $-0.05\pm 0.02$ & $-0.047 [-0.075 \ldots -0.04]$ 
    & $-0.0448\pm 0.0077$\\
\hline
\end{tabular}
\caption{Chiral predictions at LO, NLO and NNLO for the $\pi K$
scattering lengths and from Roy-Steiner (RS) equations. The 
uncertainty for the 2-loop results is taken from the various
parameter variations discussed in \cite{Bijnens:2004bu}.}
\label{tab:piK}
\vspace{-0.5cm}
\end{center}
\end{table}
is given in Tab.~\ref{tab:piK} together with the results of the
Roy-Steiner (RS) equations  \cite{Buettiker:2003pp}. The agreement
of the two-loop predictions (central values) is quite satisfactory, 
but  the precision achieved in the RS framework is not as good as
in the case of $\pi\pi$ scattering. This is largely due to the worse
and partly inconsistent data base. Similarly, the scattering lengths
extracted from these data are not very precise and fairly scattered,
see e.g. Fig.~1 in Ref.~\cite{Bernard:1990kx}. More recent measurements
in D- and B-meson decays with high statistics 
\cite{Poluektov:2004mf,Aitala:2005yh,Link:2009ng,Aubert:2008bd}
allow in principle for a better determination of $a_{1/2}$ and  $a_{3/2}$,
but none of these has performed an isospin decomposition. This is
urgently called for. In principle, lattice QCD can have some impact here.
However, the present situation as summarized for $a_{1/2}$ in
Fig.~\ref{fig:Kpi} (the date are taken from the
recent paper \cite{Lang:2012sv}, see also the references therein) is everything
but satisfactory, there is a large spread of the results. For $a_{3/2}$,
there is better agreement between the lattice results but also some
tension with the value obtained from the RS analysis. Clearly, more
work is needed to clear up the situation for this process.

\section{Lesson 3: Pion-nucleon scattering}

Let me now turn to elastic pion-nucleon scattering, $\pi N\to\pi N$, which
is the simplest (and also the most fundamental) scattering process 
involving nucleons. As in the case of
pion-kaon scattering, one has total isospin 1/2 and 3/2. Often used
are the isoscalar and isovector scattering length, $a^+$ and $a^-$,
respectively, with $a_{3/2} = a^+ - a^-$ and $a_{1/2} = a^+ + 2a^-$.
The LO prediction for the isoscalar/isovector scattering length
is quite intriguing: 
\begin{equation}
a_{\rm LO}^+ = 0~,~~ a_{\rm LO}^- =
\frac{1}{1+M_\pi/m_p}\frac{M_\pi^2}{8\pi F_\pi^2} = 79.5 \cdot 10^{-3}/M_\pi~.
\end{equation}
Much is known about the chiral corrections to these results that were
first addressed in \cite{Bernard:1993fp}. While the  chiral expansion 
for $a^-$ converges fast \cite{Bernard:1995pa},
there are large cancellations in $a^+$, so that even its sign is 
not known from scattering data, see e.g. table~3 in \cite{Fettes:2000xg}
(see also the more recent work in Ref.~\cite{Alarcon:2012kn}). 
However, there is a wonderful
alternative to get a hand on these scattering lengths, namely hadronic atoms.
These are electromagnetic bound states of two oppositely charged hadrons.
Due to the large spatial extent of these objects, the strong interactions 
lead to small perturbations in the observed level spectrum. In particular,
there is a shift of the ground state energy ($\Delta E_{1s}$) and further, 
due to channel coupling, this level acquires a width $\Gamma_{1s}$. 
Due to the small three-momenta in such a system, we are dealing essentially with 
scattering at zero energy or, stated differently, the energy shift and 
width can be expressed in terms of the corresponding scattering lengths.
There are many species of such hadronic atoms, like pionium ($\pi^+\pi^-$)
already discussed, pionic hydrogen  ($\pi^- p$), pionic deuterium  ($\pi^- d$)
and their kaonic cousins ($K^- p, K^- d$). Hadronic atoms  can be analyzed 
most systematically and precisely in suitably tailored non-relativistic
EFTs, see \cite{Gasser:2007zt} for a comprehensive review. For the case
of pion-nucleon scattering, the corresponding high-precision theoretical
framework for pionic hydrogen and pionic deuterium has been provided
in Refs.~\cite{Gasser:2002am} and \cite{Baru:2011bw}, respectively (see
also the contribution from Martin Hoferichter to these proceedings). On the
experimental side, superb experiments have been performed at PSI for
many years, culminating in precise measurements of the energy shift and
width of pionic hydrogen \cite{Gotta:2008zza} and the energy shift 
in pionic deuterium \cite{Strauch:2010vu}. 
\begin{SCfigure}[][t]
\includegraphics[width=0.40\textwidth]{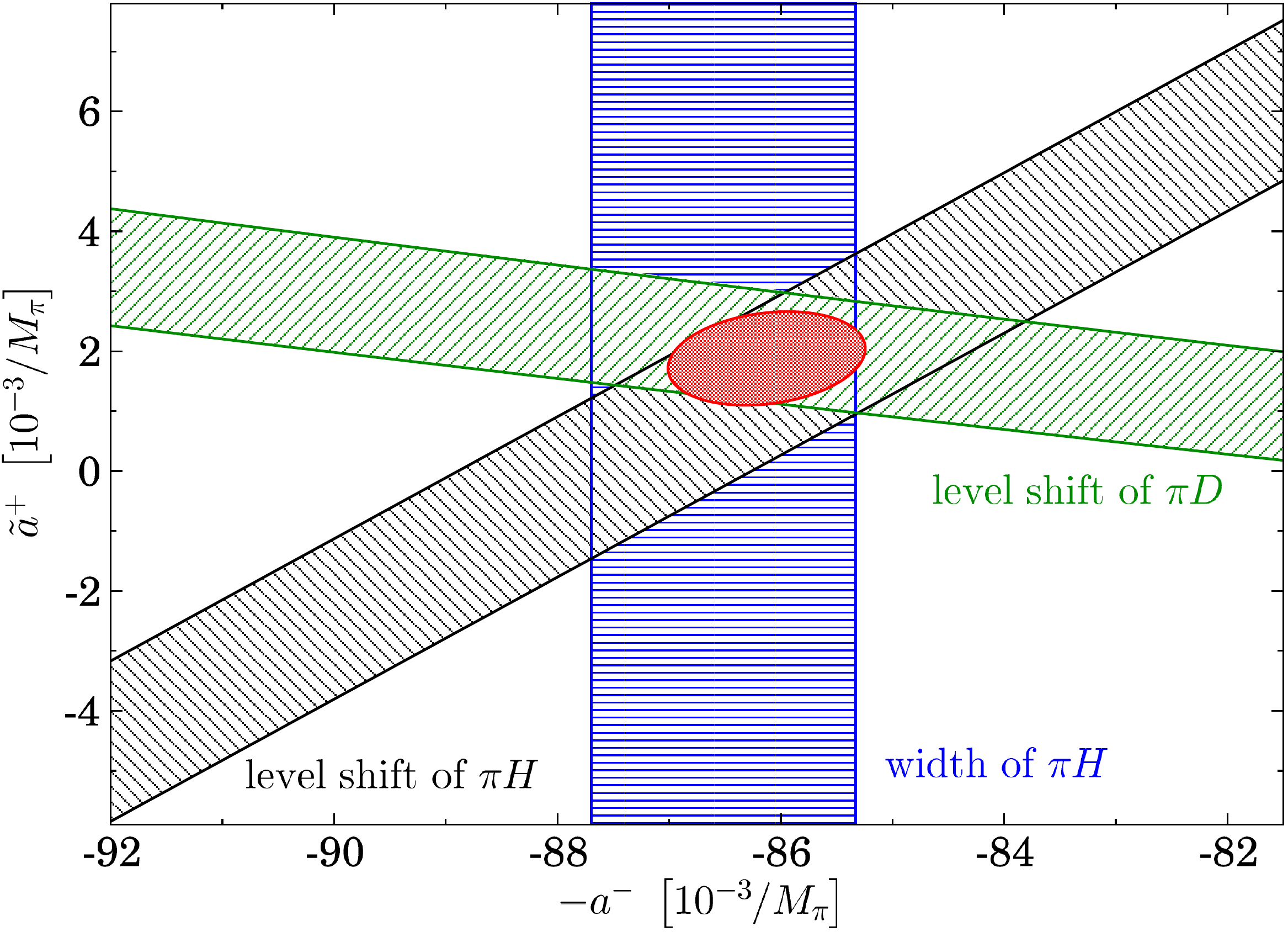}~~~~~~~
\caption{Extraction of the isoscalar and isovector
scattering lengths from pionic hydrogen and deuterium.
Here, $\tilde{a}^+ = a^+ +\frac{1}{1+M_\pi/m_p}\times $
$\left\{\frac{\Delta M_\pi^2}{\pi F_\pi}c_1 - 
2\alpha f_1\right\},$ with $\Delta M_\pi^2 = M_\pi^2 -
M_{\pi^0}^2$, $\alpha$ the fine structure constant and
$c_1$/$f_1$ are strong/electromagnetitic NLO LECs with
$c_1 \simeq 0.9\,$GeV$^{-1}$ and $|f_1| \leq 1.4$~GeV$^{-1}$.
Figure courtesy of Martin Hoferichter.
}
\label{fig:piN}
\end{SCfigure}
The analysis of these data within these EFTs leads to
a very precise extaction of the scattering lengths  \cite{Baru:2011bw}
\begin{equation}
a^+ =(7.6 \pm 3.1) \cdot 10^{-3}/M_\pi~, ~~~
a^-=(86.1 \pm 0.9) \cdot 10^{-3}/M_\pi~.
\end{equation}
Note that the value for the isovector scattering length differs by just 8\% form
the LO value, whereas the isocalar one is positive and only slightly larger in
value than various contributions to it. This underlines that {\sl only} within a
consistent framework like the employed EFT one is able to extract this value.
Also, using the GMO sum rule, the authors of \cite{Baru:2011bw} find a precise
value for the pion-nucleon coupling constant, ${g_{\pi N}^2}/(4\pi) =
13.69(12)(15)$, where the first error stems from the scattering lengths and
the second one from the integral over the $\pi^- p$ cross section. This is
consistent with other determinations from pion-nucleon scattering or from
peripheral nucleon-nucleon phase shifts.

\section{Lesson 4: Antikaon-nucleon scattering}

Next, I consider the reaction $K^-p \to K^- p$. It is a fundamental scattering 
process with strange quarks involving baryons. The dynamics of this process
is driven by channel couplings and leads to the dynamic generation of the
$\Lambda (1405)$ resonance \cite{Dalitz:1959dn,Dalitz:1960du}
that resides between the $\pi \Sigma$ and $\bar K
N$ thresholds and is certainly not a simple three-quark state. This reaction
is therefore a major playground of {\em unitarized CHPT (UCHPT)}. Due to the open
channels below the $K^-p$ threshold, the two scattering lengths $a_0$ and $a_1$
are complex-valued quantities, which means that in this case we deal with four
numbers. Before continuing,
it is important to point out the differences between CHPT and unitarized
versions thereof.   CHPT is an exact representation of the chiral Greens function of
QCD, which expands matrix elements in powers of small momenta and quark masses.
Crossing and analyticity are in general fulfilled (if a proper regularization
scheme is employed), whereas
due to the underlying power counting, unitarity is fulfilled perturbatively.
CHPT is formulated in terms of the lowest-lying hadronic degrees of freedom
and all effects from resonances are subsumed in the low-energy constants.
If one wants to describe resonances explicitely -- as it is the case here --
some resummation scheme that fulfills 2-body unitarity exactly is needed. 
There are various formalism available
to achieve that, mostly based on Lippman-Schwinger or Bethe-Salpeter equations.
In general, one gives up crossing symmetry and also some
model-dependence is induced as there is a cornucopia of unitarization schemes.
Further, the power counting is not applied to the reaction amplitude but
rather to the kernel of the scattering equation, as
it is e.g. frequently and successfully done in chiral nuclear effective field
theory. Having said that, it is also important to stress that UCHPT has been
quite successful in describing various scattering processes and the dynamic
generations of resonances like the  $\Lambda(1405)$, $S_{11}(1535)$,
$S_{11}(1650)$,  and others. 

\begin{figure}[t]
\centering
\includegraphics[width=0.44\textwidth]{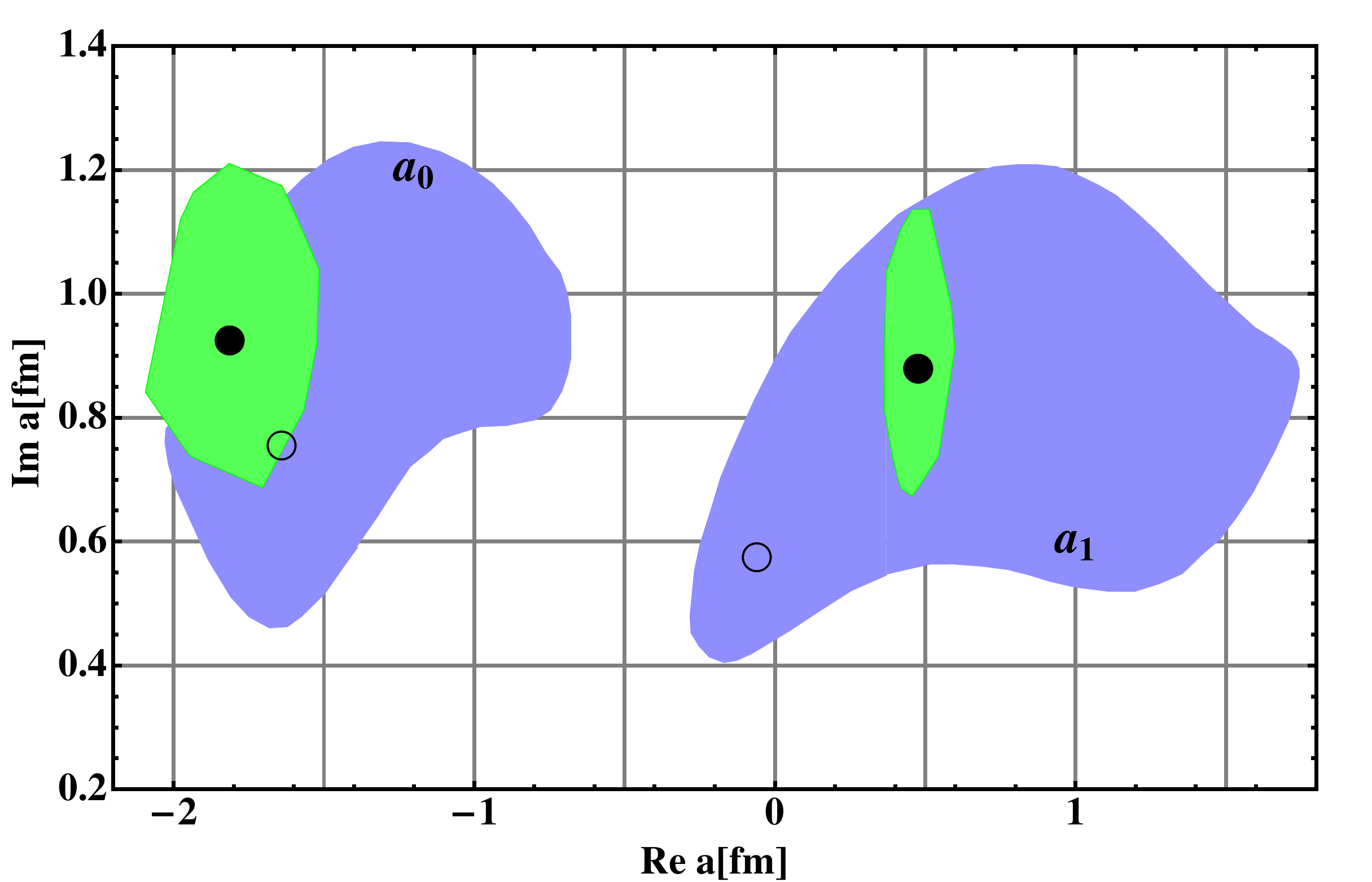}
~~~~~~~~\includegraphics[width=0.4\textwidth]{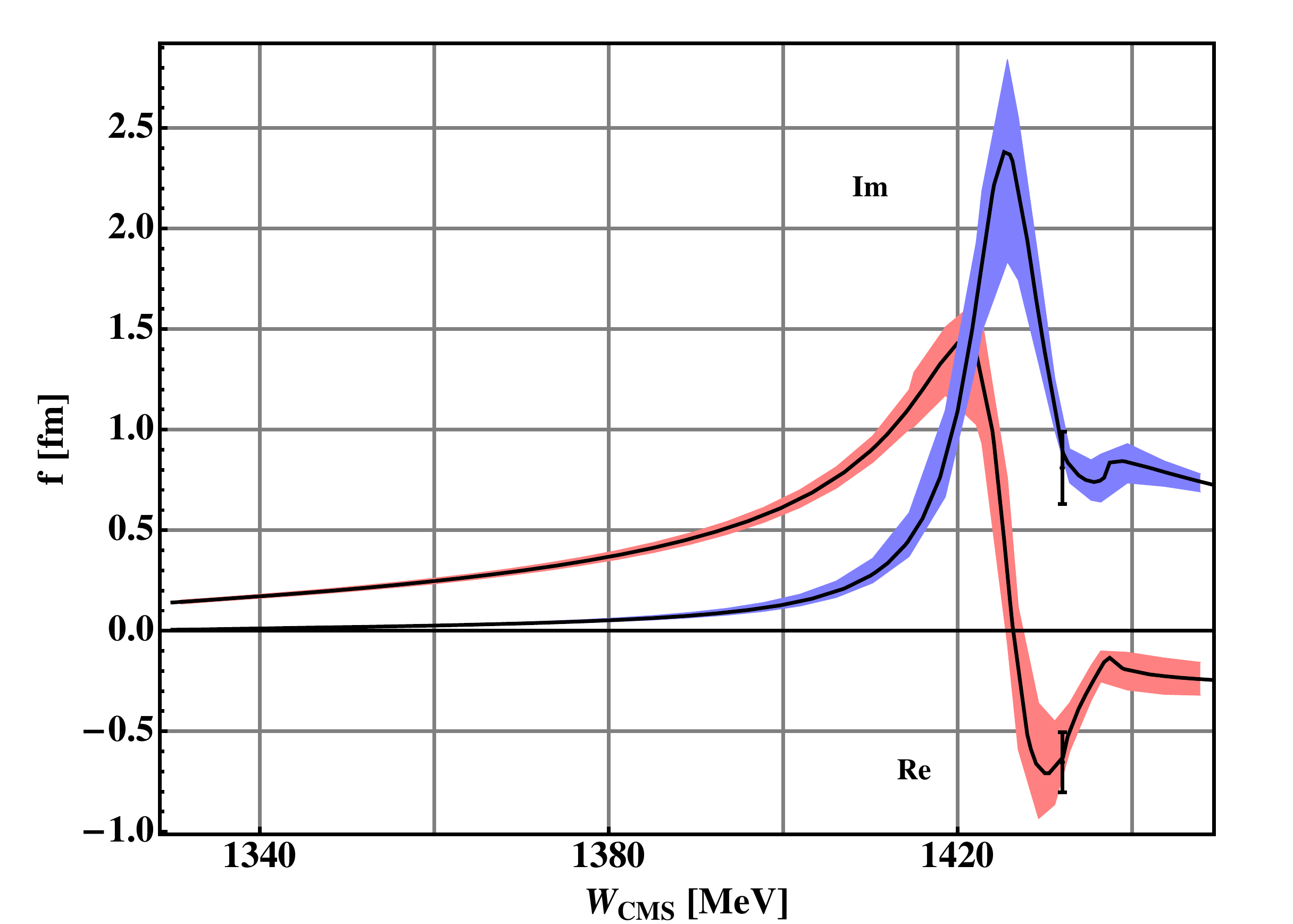}
\caption{Left panel: Real and imaginary part of the isospin $T=0$ 
    and $T=1$ ${KN\to KN}$
    scattering lengths. The light shaded (green) areas correspond to the 
    $1\sigma$ region of \cite{Mai:2012dt} around the central value (full circles). 
    The darker (blue) areas correspond to the $1\sigma$ region around central 
    value (empty circle) from Ref.~\cite{Borasoy:2006sr}.
    Right panel: Real and imaginary part of the $K^-p\to K^-p$ scattering
    amplitude. The shaded band indicates the uncertainty of the calculation. 
    The data point at $W_{\rm cms}=M_K+m_p$ is determined from the energy
    shift and width of kaonic hydrogen from the SIDDHARTA experiment.
}
\label{fig:KN}
\end{figure}

But back to $K^-p$ scattering. A long-standing puzzle in these field,
namely the inconsistency of the DEAR kaonic hydrogen data with the
scattering data (first pointed out in \cite{Meissner:2004jr} and then 
confirmed by many others)  was recently resolved by the fine experiment
of the SIDDHARTA collaboration, who measured the properties of
kaonic hydrogen with high precision, $\Delta E_{1s} = -283 \pm 36 {\rm (stat)} 
\pm 6 {\rm (syst)}$~eV and $\Gamma_{1s} = 541 \pm 89 {\rm (stat)} \pm 
22 {\rm (syst)}$~eV \cite{Bazzi:2011zj}. Based on this, the
 kaonic hydrogen and the older scattering data can now be analyzed
 consistently, using the  chiral Lagrangian at NLO, as shown by the
three groups \cite{Ikeda:2011pi,Ikeda:2012au,Mai:2012dt,Guo:2012vv}
(see also \cite{Cieply:2011nq}). Here, I report the results from 
Ref.~\cite{Mai:2012dt}. 14 LECs and 3 subtraction constants were fitted
to scattering data in the channels $K^-p \to K^-p$, $\bar K^0n$,
$\Sigma^\pm\pi^\mp$, and $\Sigma^0\pi^0$ for laboratory momenta
${p_{\rm lab} \leq 300}$~MeV together with  the SIDDHARTA data.
This allows for  good description of the antikaon-proton cross section data 
and an accurate determination of the  scattering lengths, 
\begin{equation}\label{eq:a}
a_0  = -1.81^{+0.30}_{-0.28} + i~ 0.92^{+0.29}_{-0.23}~{\rm
  fm}~, ~~~
a_1 = +0.48^{+0.12}_{-0.11} + i~ 0.87^{+0.26}_{-0.20}~{\rm fm}~.
\end{equation}
The improvement as compared to using scattering data only 
\cite{Borasoy:2006sr} is clearly visible in Fig.~\ref{fig:KN} (left panel).
These numbers are similar to the ones reported by Ikeda et al.
\cite{Ikeda:2011pi,Ikeda:2012au}. Therefore, 
these fundamental chiral SU(3) parameters have now been
determined with about an accuracy of $\sim~15\%$.
One can extrapolate the amplitudes of elastic $K^-p$
scattering to the subthreshold region, i.e. center-of-mass energies
$1330 \leq W_{\rm cms} \leq 1450$ MeV. The result is presented in the right
panel of Fig.~\ref{fig:KN}.
For both real and imaginary parts of the amplitude the maximum lies close
to the $KN$ threshold and is quite narrow, which indicates the presence of a close-by pole. 
It is also worth mentioning that the error band gets smaller for lower
energies, different to the recent analysis by Ikeda et al.~\cite{Ikeda:2011pi,Ikeda:2012au}.
We note that although Ikeda et al. and  \cite{Mai:2012dt} describe the scattering and bound
state data equally well, the subthreshold amplitude is very different. This is
presumably due to the different approximation made in these two approaches.
Therefore, a truly model-independent determination of this subthreshold
amplitude is not yet available.

\section{Lesson 5: Goldstone boson scattering off {\boldmath$D$} and
{\boldmath$D^\star$}-mesons}

Finally, let us consider scattering the Goldstone boson octet $(\pi, K ,\eta)$  
off the $D$-meson triplet $(D^0, D^+, D_s^+)$. This involves the
positive-parity  scalar charm-strange  meson $D_{s0}^*(2317)$ which
has a very narrow, isospin-violating width and is interpreted by 
some groups as a molecular $DK$ state. Here, we are mostly interested 
in the scattering length in the channel with $(S,I)=(1,0)$, as its 
value can tell us something about the possible nature of the scalar 
meson. As I will show, this field enjoys a  healthy interplay of 
lattice QCD and UCHPT. But a few general remarks on the calculation
of the scattering process $\phi D \to\phi D$ are in order. This is 
an interesting problem, as it involves a variety of scales and is 
also multi-faceted. First, light particles related to the chiral 
symmetry of QCD are involved, thus we can perform a chiral expansion 
in momenta and quark masses. Second, as the $D$-meson contain charm 
quarks, we can exploit heavy quark symmetry and perform an expansion 
in $1/m_c$. Third, one must consistently include isospin-violation
effects. These are on one hand generated by the strong interactions 
($m_d \neq m_u$) and on the other hand of electromagnetic origin ($q_u \neq q_d$).
In total, one has to consider 16 channels with different total strangeness and
isospin. Some of these are perturbative, but most are non-perturbative and
require resummation, which can lead to the dynamical generation of
molecules. 

Before addressing the actual calculations, let me discuss the relation
between the scattering length $a$ and the nature of the state under
consideration, see Refs.~\cite{Weinberg:1965zz,Baru:2003qq}
for a derivation:
\begin{equation}
 a = -2 \left( \frac{1-Z}{2-Z} \right) \frac1{\sqrt{2\mu\epsilon}} \left(
 1+\mathcal{O}(\sqrt{2\mu\epsilon}/\beta) \right),
 \label{eq:weinberg}
\end{equation}
where $\mu$ and $\epsilon$ are the reduced mass of the two-hadron 
system and the binding energy, respectively, and $Z$ is the wave function
renormalization constant ($0 \leq Z\leq 1$).
 Corrections of the above equation come from neglecting the range of
forces, $1/\beta$, which contains information of the $D_s\eta$ channel.
Were the $D_{s0}^*(2317)$ a pure $DK$ bound state ($Z=0$), the value of 
the $DK(I=0)$ scattering length would be $a=-1.05$~fm.
\begin{figure}[t]
\centering
\includegraphics[width=0.84\textwidth]{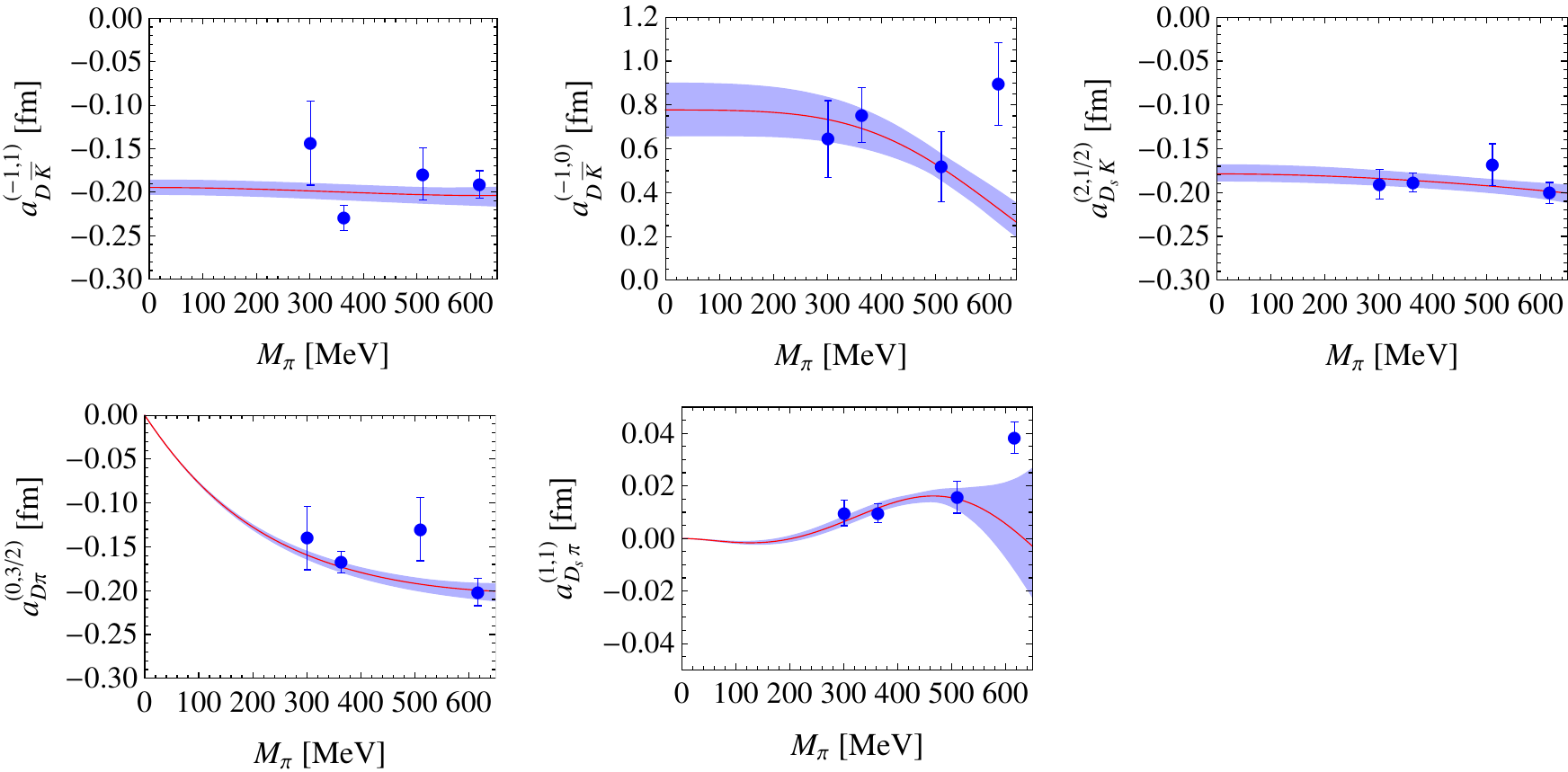}~~~~~
\caption{Fit to lattice data for various channels 
with 5 parameters. The subtraction constant is 
determined from fixing the pole in the $(S,I)=(1,0)$ channel to 
2317.8~MeV. The points at the highest pion mass in each channel are not fitted.
}
\label{fig:Dpi}
\end{figure}
In \cite{Liu:2012zya}, new lattice data using the MILC plus Fermilab actions
for the channels $D\bar K (-1,1), (-1,0)$, $D_sK(2,1/2)$, $D\pi(0,3/2)$ and
$D\pi(1,1)$ were analyzed based on the UCHPT formalism developed in
\cite{Guo:2009ct} (for related work, see 
\cite{Liu:2009uz,Geng:2010vw,Cleven:2010aw,Wang:2012bu}).
These are more data than previously available allowing 
in particular for the inclusion of $N_c$-suppressed operators
of the NLO effective Lagrangian (there are 5 LECs at this
order from which 3 are formally subleading in $1/N_c$ and one subtraction
constant).
If one requires the $D_{s0}^* (2317)$ to be a $DK$-molecule by a proper
choice of the subtraction constant (i.e. having 
a pole at the proper mass in this channel), 
one has 5 fit parameters that describe the lattice data well, 
see Fig.~\ref{fig:Dpi}. In this fit, all parameters come out of natural 
size and the  large-$N_c$ hierarchy obeyed. The scattering length 
in the $DK(I=0)$ channel comes out as
$a(DK(I=0)) = -0.85^{+0.07}_{-0.05}~\mbox{fm}$, which is consistent
with the molecular interpretation (for a more detailed discussion,
also concerning a different fit procedure, see \cite{Liu:2012zya}).
Having pinned down all the LECs, one finds an improved prediction 
for the isospin-violating width 
$\Gamma(D_{s0}^* (2317)^+ \to D_s^+ \pi^0) = (89\pm 27)\,{\rm keV}\,$
(for earlier work in the molecular picture, see 
\cite{Faessler:2007gv,Lutz:2007sk,Guo:2008gp}).
This is very different from typical quark model predictions,
which find this width to be about a few keV \cite{Godfrey:2003kg},
and thus an accurate  measurement of this quantity is called for.
I wish to end with a few remarks on the $D_{s1}(2460)$. As
$M_{D_{s1}(2460)}{-}M_{D_{s0}^*(2317)} \simeq M_{D^*}{-}M_D$, it is most
likely a $D^\star K$ molecule (if the $D^\star_{s0} (2317)$
is a $DK$ molecule).  Goldstone boson scattering off $D$- 
and $D^\star$-mesons was considered in \cite{Cleven:2010aw}.
The most interesting observation made concerns the
mass and binding energy of a molecular state of a heavy meson $(H)$
and a kaon. As its mass is given by $M_{\rm mol} = M_K + M_H - \epsilon$,
the mass should depend linearly on the kaon mass, very different
from a genuine multi-quark state, whose mass depends linearly on the
strange quark and thus quadratically on the kaon mass. This behaviour
can, of course, be investigated on the lattice.

\section{Short summary \& outlook}

As I have shown, there has been much progress in our understanding of
hadron-hadron scattering lengths since Weinberg's seminal paper in 1966.
Most advanced is theory and experiment in the case of pion-pion scattering.
Here, lattice QCD still has to provide an acccurate number for $a_0$.
Matters are much less satisfactory for pion-kaon scattering - as both 
experiment and lattice QCD have to deliver more precise values for 
both scattering lengths. A measurement of the properties of $\pi K$ atoms
would certainly be very welcome \cite{Adeva:2009zz,Schacher}.
Very different to that, the combination of
chiral EFTs with precision data has led to a high precision determination
of the pion-nucleon scattering lengths $a^+$ and $a^-$. This should 
challenge lattice practioners to provide a similarly accurate ab initio
calculation. In case of antikaon-nucleon scattering, there has been
recent progress by analysing scattering data and kaonic hydrogen data
provided by SIDDHARTA and the corresponding scattering lengths are now
known with an accuracy of about 15\%. Here, a measurement of kaonic deuterium
would provide further stringent constraints 
\cite{Doring:2011xc,Shevchenko:2012np,Faber:2010iw}. Finally,  including 
charm quarks, the interplay of unitarized CHPT and lattice QCD complemented
by experiment can deepen our understanding of heavy-light systems, thus
providing a bridge from chiral dynamics to the physics of heavy quark flavors.

\section*{Acknowledgements}

I thank all my collaborators for sharing their insight into the
topics discussed here.


\end{document}